\documentclass[10pt,conference]{IEEEtran}
\IEEEoverridecommandlockouts
\usepackage{cite}
\usepackage{arydshln}
\usepackage{amsmath,amssymb,amsfonts}
\usepackage{algorithmic}
\usepackage{graphicx}
\usepackage{textcomp}
\usepackage{subcaption}
\usepackage[dvipsnames]{xcolor}

\usepackage{array}
\newcolumntype{C}[1]{>{\centering\arraybackslash}p{#1}}

\usepackage[utf8]{inputenc}
\usepackage{url}
\usepackage{hyperref}
\usepackage{tabularx}
\usepackage{caption}
\usepackage{xcolor,colortbl} 
\usepackage{flushend}
\usepackage{enumitem}
\usepackage{tcolorbox}

\usepackage{amsmath,amssymb,amsfonts}
\usepackage{algorithmic}
\usepackage{graphicx}
\usepackage{textcomp}

\usepackage{multirow}
\usepackage{makecell}
\usepackage{listings}

\usepackage{tablefootnote}
\usepackage{threeparttable}

\newcommand{\etal}{\textit{et al.}}

\newcommand{\circled}[1]{\raisebox{.5pt}{\textcircled{\raisebox{-.9pt} {#1}}}}

\definecolor{comment_green}{RGB}{133,170,130}

\DeclareRobustCommand{\summarybox}[1]{%
\begin{tcolorbox}[
    boxrule=.5pt,
    left=3pt,
    right=3pt,
    top=3pt,
    bottom=3pt,
    colback=black!5!white,
    colframe=black!75!black
    ]
    #1
\end{tcolorbox}
}
\definecolor{LightGray}{gray}{0.9}

\begin{document}

\title{Leveraging multi-task learning to improve the detection of SATD and vulnerability}

\author{
    \IEEEauthorblockN{Barbara Russo}
        \IEEEauthorblockA{\textit{Free University of Bozen-Bolzano}\\
        Bolzano, Italy\\
        barbara.russo@unibz.it
        }
    \and
    \IEEEauthorblockN{Jorge Melegati}
        \IEEEauthorblockA{\textit{Free University of Bozen-Bolzano}\\
        Bolzano, Italy\\
        jorge.melegati@unibz.it
    \and
    \IEEEauthorblockN{Moritz Mock}
        \IEEEauthorblockA{\textit{Free University of Bozen-Bolzano}\\
        Bolzano, Italy\\
        moritz.mock@student.unibz.it
        }
    }
}

\maketitle

\begin{abstract}
Multi-task learning is a paradigm that leverages information from related tasks to improve the performance of machine learning. Self-Admitted Technical Debt (SATD) are comments in the code that indicate not-quite-right code introduced for short-term needs, i.e., technical debt (TD). Previous research has provided evidence of a possible relationship between SATD and the existence of vulnerabilities in the code.
In this work, we investigate if multi-task learning could leverage the information shared between SATD and vulnerabilities to improve the automatic detection of these issues.
To this aim, we implemented VulSATD, a deep learner that detects vulnerable and SATD code based on CodeBERT, a pre-trained transformers model. We evaluated VulSATD on MADE-WIC, a fused dataset of functions annotated for TD (through SATD) and vulnerability. We compared the results using single and multi-task approaches, obtaining no significant differences even after employing a weighted loss.
Our findings indicate the need for further investigation into the relationship between these two aspects of low-quality code. Specifically, it is possible that only a subset of technical debt is directly associated with security concerns. Therefore, the relationship between different types of technical debt and software vulnerabilities deserves future exploration and a deeper understanding.
\end{abstract}


\section{Introduction}
\label{sec:introdution}

Low-quality code is a crucial problem in maintenance. It is typically harder to understand, debug, and modify. Poorly written code often lacks clarity, proper documentation, and structure, which makes identifying and fixing issues more time-consuming for developers. This increased effort leads to higher maintenance costs over the software's lifecycle. When this happens, the portion of the code of low-quality is called Technical Debt (TD)~\cite{Fowler1999,Cunningham2009}.  
Low-quality code can also increase the risk of vulnerabilities being introduced and make identifying and fixing these flaws significantly more challenging.  
Detecting and predicting TD and vulnerability in code is crucial for modern software engineering. Numerous studies have independently employed deep learning techniques to address either of the two facets of low-quality code. However, only a few recent works, such as those by Izurieta \etal~\cite{Izurieta2018}, Russo et al.\cite{Russo2022}, Ferreyra et al.\cite{FerreyraEtAl2024}, Edbert \etal~\cite{Edbert2023}, have hypothesized a potential relationship between TD and vulnerability. In this work, we aim to explore further and better understand the connection between these two forms of low-quality code. To this end, we employ multi-tasking learning. Multi-task learning is a paradigm that leverages information of related tasks to improve the performance of machine learning~\cite{Zhang2022}. Recently, this approach has also been applied to research problems in software engineering, such as predicting issue priorities with issue categories~\cite{Li2022} and the type and value of the tokens in code simultaneously~\cite{Liu2022}. 
The input we pass to the multi-task model consists of pairs of comments and code functions. 
Comments may contain the information left by developers to identify TD. A comment in which developers acknowledge a code to be TD is called Self-Admitted Technical Debt (SATD) ~\cite{Russo2022, Bavota2016, Potdar2014, Maldonado2015}. Such comments can be automatically detected, and the associated code can be identified and removed or modified to mitigate the debt (paying back the technical debt)~\cite{Ren2019TOSEM, Guo2021, Yu2022TSE}. Code functions can be annotated as vulnerable and 
 automatically detected ~\cite{ZhouEtAl2019,Chakraborty2022, SteenhoekEtAl2023,FuTantithamthavorn2022}. Russo \etal~\cite{Russo2022} detected more than 55\% of the Java files of the Chromium project with both TD and vulnerability. 
Therefore, our research goal is \textit{to compare multi-task with single-task learning of TD and vulnerable functions to understand the relation between these two facets of low-quality code.} In other terms, we aim to see whether the information about SATD and vulnerability can improve the detection of one or the other aspect of a function. 
To this end, we propose \textit{VulSATD}, a deep learning approach designed to detect Self-Admitted Technical Debt (SATD) and/or vulnerable functions written in C. VulSATD leverages state-of-the-art (SOTA) natural language processing (NLP) tools: Byte Pair Encoding (BPE) for tokenization and CodeBERT~\cite{Feng2020}, a bimodal pre-trained model for text and code embeddings. CodeBERT has previously been successfully applied to both vulnerability detection~\cite{Fu2022} and SATD detection~\cite{Aiken2023}. VulSATD extends this capability by classifying SATD and/or vulnerable code using different architectures in its final layers, supporting both multi-task and single-task learning paradigms. VulSATD has been evaluated on MADE-WIC~\cite{MockEtAl2024Dataset}, a recently published fused dataset combining two publicly available SOTA vulnerability datasets, Devign~\cite{ZhouEtAl2019} and Big-Vul~\cite{Fan2020}, alongside data from three major open-source projects: Chromium, the Linux Kernel, and Mozilla Firefox. MADE-WIC includes annotations for SATD—based on the MAT tag annotation~\cite{Guo2021} and the patterns proposed by Potdar and Shihab~\cite{Potdar2014}—as well as vulnerability annotations derived from the original datasets and security-related concerns. This non-synthetic dataset enables more realistic and accurate solutions to the classification problem~\cite{YangEtAl2023}. Given the inherent class imbalance in the dataset, VulSATD has been further customized with a weighted loss function to mitigate the effects of bias. 

Our results are negative. We applied both the multi-task and single-task versions of VulSATD, with and without class balancing. In all our experiments, we did not observe any significant improvement in model performance across the various datasets within MADE-WIC. Furthermore, any observed variations in performance, whether increases or decreases, were minimal and inconsistent.

In summary, this article makes the following contributions:
\begin{itemize}
\item We introduce VulSATD, a deep learning-based approach designed to detect SATD and vulnerable code in C, leveraging advanced NLP techniques such as Byte Pair Encoding (BPE) and CodeBERT.
\item We evaluate both multi-task and single-task learning paradigms within VulSATD, providing insights into their effectiveness in classifying SATD and vulnerable functions.
\item
VulSATD is applied to MADE-WIC, a recent dataset that integrates data from SOTA vulnerability datasets and annotations for SATD, and vulnerable functions from real-world projects.
\item
We investigate the impact of class imbalance and apply weighted loss functions to address this issue, highlighting the limited and inconsistent improvements in performance across different settings.
\end{itemize}
Despite the lack of significant improvements, our findings provide valuable insights into the challenges of simultaneously addressing SATD and vulnerability, emphasizing the need for further research by, for example, investigating the mutual relation by type of SATD and vulnerability.

The rest of this article is organized as follows: Section~\ref{sec:motivation} discusses the motivation of the work. Section \ref{sec:approach} overviews the methodology of the work and the implementation details together with reference to the replication package. Section~\ref{sec:evaluation} introduces the research questions while Section~\ref{sec:datasets} describes the extension and annotation of the datasets. Section \ref{sec:results} reports the experimental results and, in Section~\ref{sec:discussion}, we discuss their implications and possible future work. In Section~\ref{sec:relatedWork}, we summarize relevant literature in terms of SATD and vulnerability detection. Section~\ref{sec:threats} discusses the threats to validity whereas Section~\ref{sec:conclusions} reflect of the results and the future development of our work.

\section{Motivation}
\label{sec:motivation}
 
Listing~\ref{lst:example} presents an example of code containing both SATD comments and vulnerabilities. The comment `FIXME' highlights that the size of \texttt{sigmask} is specific (an even multiple of the size of a long integer), while the function at line 21 copies memory from \texttt{set} to \texttt{sigmask} without controlling the size. This situation is a typical case that can cause a buffer overflow, which could be exploited. The developer is aware of the issue but leaves it for future maintenance. However, they may not realize that the time required to fix the problem could be very long~\cite{Bavota2016}, leaving the vulnerability exposed for an extended period.
\begin{lstlisting}[
    basicstyle=\footnotesize,
    escapechar=!, 
    language=C, 
keywordstyle=\color{blue}, 
    breaklines=true, 
    numbers=left,
    label={lst:example},
    caption={Example of vulnerable code containing a SATD comment.},
    xleftmargin=2em,
    framexleftmargin=1.5em,
    tabsize=1,
    numbersep=3pt,
    columns=fullflexible,
    float=t
]
/* FIXME: this code assumes that sigmask is an even multiple of the size of a long integer. */
unsigned long *src = (unsigned long const *) set;
unsigned long *dest = (unsigned long *) &( thread.p->sigmask);

switch (how) { 
    case SIG_BLOCK:
    for (i = 0; i < (sizeof (sigset_t) / sizeof (unsigned long)); i++)
    {
        /* OR the bit field longword -wise. */
        *dest++ |= *src++;
    }
    break;
    case SIG_UNBLOCK:
    for (i = 0; i < (sizeof (sigset_t) / sizeof (unsigned long)); i++)
    {
        /* XOR the bitfield longword -wise. */
        *dest++ ^= *src++;
    }
    case SIG_SETMASK:
    /* Replace the whole sigmask. */
    memcpy (&( thread.p->sigmask), set , sizeof (sigset_t));
    break; 
}
\end{lstlisting} 
Another important point is how pervasive is this phenomenon. We perform a frequency test on the larger and heterogenous portion of MADE-WIC (the Big-Vul dataset). Table~\ref{tab:contingency_big_vul} illustrates its contingency table for SATD and vulnerable functions. The Chi-Square test rejects the null hypothesis that a function's vulnerability is independent of its status as TD ($\chi ^2$ = 2586.6) and p-value=0.0). These findings suggest a form of informational dependency between the two facets of low-quality code. In this work, we aim to explore this relationship further using multi-task learning.  
\begin{table}[!ht]
\renewcommand{\arraystretch}{1.2}
\centering
\caption{Contingency table of vulnerable and SATD in MADE-WIC/Big-Vul.}
\label{tab:contingency_big_vul}
\begin{tabular}{lccc}
\hline
\textbf{}& \textbf{Non-vulnerable} & \textbf{Vulnerable}\\ \hline
Non-SATD & 134,515 & 7,791 \\
SATD & 1,395 & 657 \\\hline
\end{tabular}
\end{table}


\section{Dataset}
\label{sec:datasets}

Our approach requires a dataset of pairs (comment, function) that are annotated as SATD (comment) and vulnerable (function). 
\begin{table*}[!ht]
    \renewcommand{\arraystretch}{1.2}
    \centering
    \caption{
    Relevant public data sets of non-synthetic data of functions annotated for vulnerability or SATD. 
    }
    \label{tab:SOTA_datasets}
    \begin{tabularx}{\linewidth}{llXrl}
         \hline
    \textbf{Name}&\textbf{Study}&\textbf{Data source}& \textbf{$\#$ functions}&\textbf{Annotation}  \\
         \hline
         &Russell \etal~\cite{Russell2019}&SATE IV, Github and Debian  & $\sim$1.3M&Vul. \\ 
         \hline
         Big-Vul&Fan \etal~\cite{Fan2020} & 348 Github projects &$\sim$265k &Vul. \\ 
          \hline
          &Harer \etal~\cite{HarerEtAl2018}& Debian Linux distribution and Github&$\sim$981k  &SATD\\ 
         \hline
        Devign &Zhou \etal~\cite{ZhouEtAl2019}& Linux kernel, QEMU, Wireshark, FFmpeg& $\sim$49k & Vul. \\ 
          \hline
          &\makecell[l]{Li \etal~\cite{LiEtAl2021} and\\ Lin \etal~\cite{Lin2020}}& 
FFmpeg, LibTIFF, LibPNG, Pidgin, VLC media player, Asterisk, HTTPD, OpenSSL, and Xen&$\sim$61k &Vul. \\
 \hline
         MADE-WIC&Mock \etal~\cite{MockEtAl2024Dataset}&Chromium, Linux Kernel, Mozilla FireFox&$\sim$688K&Vul. and SATD\\ 
         \hline ReVeal&Chakraborty \etal~\cite{Chakraborty2022} & FFmpeg, Qemu, Chrome, and Debian & $\sim$18k & SATD \\ 
         \hline
         10 Java Projects&Maldonado \etal~\cite{Maldonado2017} & Ant, ArgoUML, Columba, EMF, Hibernate, JEdit, JFreeChart, JMeter, JRuby, SQuirrel& $\sim$33k & SATD\\ 
         \hline
         20 Java Projects&Guo \etal~\cite{Guo2021} & Maldonado \etal~\cite{Maldonado2015} + Dubbo, Gradle, Groovy, Hive, Maven, Poi, SpringFramework, Storm, Tomcat, Zookeeper& $\sim$81k & SATD\\
         \hline
    \end{tabularx}
\end{table*}
We  explored the existing literature in vulnerability and SATD detection to search for non-synthetic datasets in which functions are annotated as vulnerable and/or SATD.  Table~\ref{tab:SOTA_datasets} reports the datasets analysed. 
Among them, we found  only  MADE-WIC~\cite{MockEtAl2024Dataset} whose functions are annotated both as vulnerable and TD through SATD comments. MADE-WIC fuses two datasets of functions annotated for vulnerability, Big-Vul~\cite{Fan2020} and Devign~\cite{ZhouEtAl2019}, with three open-source projects (OSPR)- Chromium, Linux Kernel, and Mozilla FireFox. The result is a dataset whose entries are annotated for SATD in two different ways - one  with the patterns of Potdar and Shihab~\cite{Potdar2014} and the other with the patterns of Guo \etal~\cite{Guo2021} and, for vulnerability, in three different ways depending on the original subset, i.e., the ones of Big-Vul~\cite{Fan2020}, Devign~\cite{ZhouEtAl2019} and WeakSATD~\cite{Russo2022}. 
Big-Vul uses references from the CVE repository to the Github repositories and the relevant commits that fix vulnerabilities. Vulnerable functions are detected in the change set of such fixing commits. 
Functions in the Devign dataset are annotated as vulnerable by first identifying potentially vulnerable commits. Vulnerable commits are identified by keywords in their messages, e.g., ``illegal'', ``leak'', and many others, and validated by manual inspection. Finally, functions are annotated as vulnerable if they are in the change set of such vulnerable commits. 
For OSPR, a function is annotated as vulnerable if it contains a weak code snippet matching the code examples of the CWE repository~\cite{Russo2022}. 
The SATD annotation of the Potdar and Shihab  \cite{Potdar2014} uses 62 different patterns in comments to annotate a comment as SATD and its related function as TD, whereas Guo \etal~\cite{Guo2021} propose the Matches task Annotation Tags (MAT) that leverages the four task annotation tags that are typically recommended by integrated development environments (IDEs): ``TODO'', ``FIXME'', ``XXX'', and ``HACK''. 
The advantage of MADE-WIC lies in its preprocessing through data fusion~\cite{BleiholderNaumann2009DataFusion}, which standardizes the datasets under a unified schema, thereby facilitating seamless integration and interchangeability of subsets during experiments.
Besides that, MADE-WIC \cite{MockEtAl2024Dataset} contains well-known datasets - the two publicly available projects of Devign and the ten largest ones of Big-Vul, which account for up to 75 per cent of the total size of the original dataset. MADE-WIC also includes the \textit{leading comment of a function} \cite{MockEtAl2024Dataset}, i.e., the source code comment just before the function code, that together with the code comments inside a function represent all comments related to the function. We will show that including the leading comment in the set of comments related to a function does impact on the SATD classification performance. 
Table~\ref{tab:datasets} summarizes the datasets of MADE-WIC.  
\begin{table}[h!t]
\renewcommand{\arraystretch}{1.2}
    \centering
 
    \caption{Datasets of  MADE-WIC  we considered in the study.}
\label{tab:datasets}
    \begin{tabular}{lp{0.8\linewidth}}
        \hline
        \textbf{Dataset} &\textbf{Projects} \\
        \hline
    OSPR&Chromium, Firefox, Linux Kernel \\
    Devign&QEMU and FFmpeg \\
        Big-Vul& Chromium, Linux Kernel, Android, PHP interpreter, FFmpeg, ImageMagick, Radare2, Kerberos v5, and Tcpdump  \\
        \hline
\end{tabular}
\end{table}
Table~\ref{tab:statistics} summarizes statistics on MADE-WIC. 
It is worth noting that they have a higher percentage of SATD instances than what is described in the literature (e.g., Bavota and Russo \cite{Bavota2016}). The percentage of vulnerable functions varies in the three datasets. The difference is mainly due to the different strategies of annotation~\ref{tab:datasets} and enable us to analyse their impact on model performance. 

\begin{table}[ht]
\renewcommand{\arraystretch}{1.2}
\centering
\caption{Demographic of our datasets.}
\label{tab:statistics}
\begin{tabular}{lrrr}
\hline
\textbf{Name} &  \textbf{\makecell{\# functions}} &  \textbf{\makecell{\# SATD functions}} & \textbf{\makecell{\# Vulnerable functions}}  \\
\hline
OSPR  & 688,134 & 9,388 (1.36\%) & 219,625 (31.9\%)  \\
Devign  & 27,282  & 2,744 (9.94\%) & 12,437 (45.5\%)  \\
Big-Vul  & 144,358 & 2,052 (1.42\%) & 8,448 (5.85\%)  \\
\hline
\end{tabular}   
\end{table}

\section{Methodology}
\label{sec:approach}
\begin{figure}[b]
    \centering
\includegraphics[width=\columnwidth]{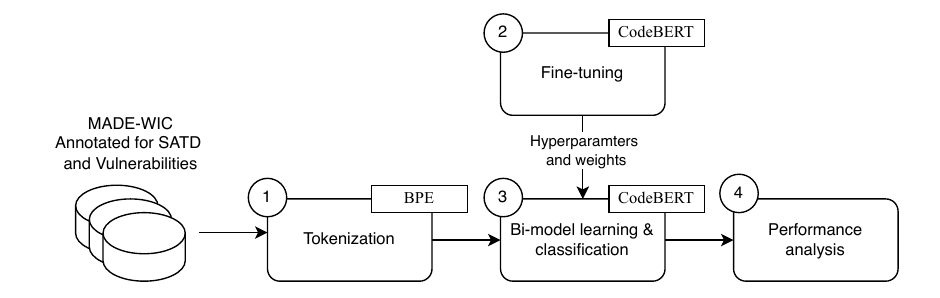}
\caption{The VulSATD approach.}
\label{fig:overview}
\end{figure}

In this section, we overview VulSATD, our approach to detect functions that are vulnerable and SATD. VulSATD comprises four major steps: \circled{1} tokenization of the input leveraging Byte Pair Encoding (BPE)~\cite{SennrichEtAl2016}, \circled{2} VulSATD fine-tuning, \circled{3} VulSATD learning and classification, and \circled{4} performance analysis, Fig.~\ref{fig:overview}. 
As we mentioned, the VulSATD architecture occurs in two different fashions: the single-task and the multi-task as described in the following.
\subsection{VulSATD architecture}\label{VulSATD}
\begin{figure*}[t]
    \centering
\begin{subfigure}[t]{0.45\textwidth}
\includegraphics[width=\columnwidth]{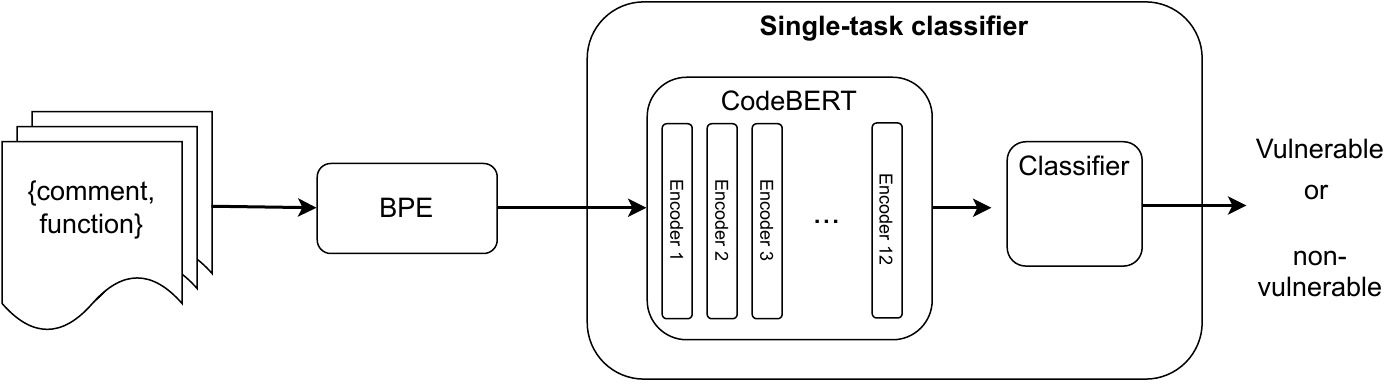}
    \caption{With single-task classifier for vulnerability (equally for SATD).}
    \label{fig:singleTaskArch}
\end{subfigure} 
\hspace{30pt}
\begin{subfigure}[t]{0.45\textwidth}
\includegraphics[width=\columnwidth]{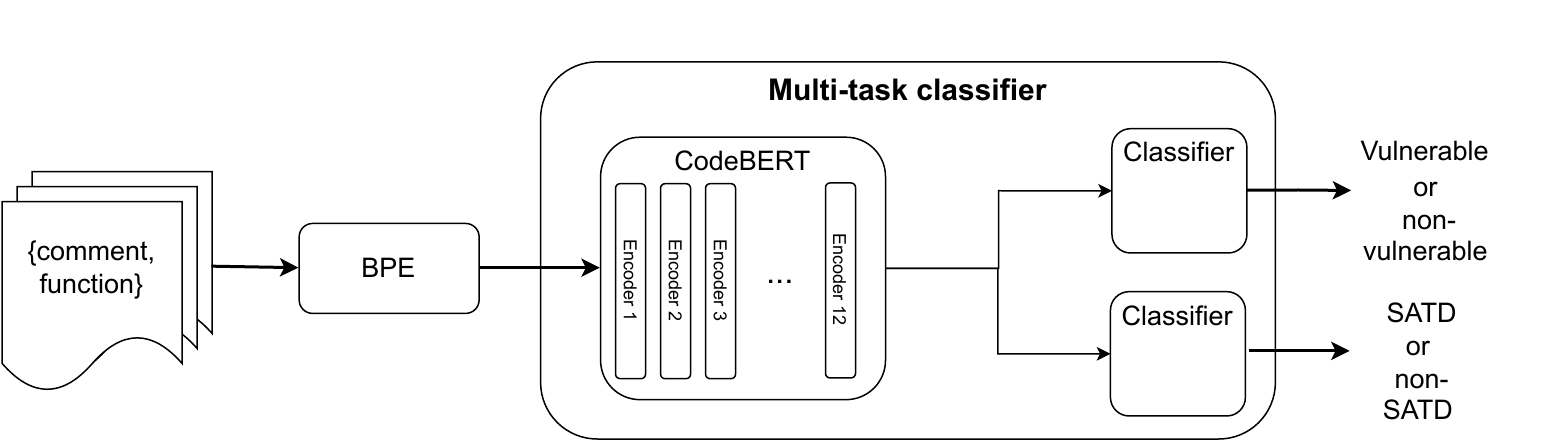}
    \caption{With multi-task classifier for vulnerability and SATD}
\label{fig:multiTasksArch}
\end{subfigure}
\caption{VulSATD architectures}
\end{figure*}
We leverage the BERT architecture with self-attention layers as it is capable of capturing long-term dependencies within a long sequence using dot-product operations and the relationship between tokens~\cite{Zhou2021, YangEtAl2023}. 
In particular, we use the CodeBERT pre-trained language model to generate a vector representation of source code~\cite{Feng2020}. CodeBERT was pre-trained on 20GB of code corpus (i.e., CodeSearchNet) using a Robustly Optimized BERT pre-training approach, ROBERTA~\cite{Liu2019}, on six programming languages (Python, Java, JavaScript, PHP, Ruby, Go) and evaluated on the C++ language thereafter.
CodeBERT is a transformer encoder with an architecture consisting of 12 layers, 768 hidden size, 12 self-attention heads, and 125 million parameters. We used a version of the model available in HuggingFace\footnote{https://huggingface.co/microsoft/codebert-base}.
Using CodeBERT helps mitigate the problem of learning with deep learners generally trained on text that may learn irrelevant features of the code~\cite{Chakraborty2022}.
The input of CodeBERT is a concatenation of two segments with a special separator token, namely \texttt{[CLS], w\textsubscript{1}, w\textsubscript{2}, ..., w\textsubscript{n}, [SEP], c\textsubscript{1}, c\textsubscript{2}, ..., c\textsubscript{m}, [EOS]}. One segment is the comment as natural language text, and another is the function code. \texttt{[CLS]} is a special token in front of the two segments, whose final hidden representation is considered as the aggregated sequence representation for classification or ranking.  
The output of CodeBERT includes the contextual vector representation of each token for both comment and function and the representation of \texttt{[CLS]}, which works as the aggregated sequence representation~\cite{Feng2020}.

A final classifier is added as last layer. The classification is performed for vulnerability and SATD (multi-task), and for vulnerability only and SATD only (single-task), Figs.~\ref{fig:singleTaskArch} and \ref{fig:multiTasksArch}. 
Both configurations learn on the $\{$comment, function$\}$ pairs of each dataset. The multi-task architecture classifies functions on each of the two tasks: SATD and vulnerability. The multi-task learner shares the layers of CodeBERT while classifying vulnerability and SATD separately. 2) A single-task version for each of the two tasks. In all cases, CodeBERT is set to bimodal with input $\{$comment, function$\}$.
The variants change the way the final classification is performed (multi-task, single-task).

\subsection{CodeBERT input}\label{sec:codeBERTInput}
\subsubsection{Comments} We apply CodeBERT in a bi-modal fashion that separates the input into comments (first mode) and functions (second mode).
As MADE-WIC provides the function code, including its internal comments and its leading comments separately, we want to understand whether to keep the comments in the function's body within the function or move them out and aggregate them with the leading comment. This will input CodeBERT with text and code separately.
Given the limited size of the CodeBERT input (512 tokens), leaving comments or vulnerabilities out in some cases.      
To evaluate this aspect, we analyse the performance of VulSATD with two types of input: when comments (leading and internal comments) are all aggregate (\textit{out}) or when the internal comments are left as originally in the body of the function (\textit{in}). 
\begin{table}[b]
    \caption{Precision, Recall, and F1 for single-task SATD and vulnerability, for different input combinations, and a fixed loss.}
    \label{tab:ST_comments_in_vs_out}
    \renewcommand{\arraystretch}{1.2}
    \centering
    \begin{tabular}{cl|C{0.06\textwidth}C{0.06\textwidth}C{0.06\textwidth}|C{0.06\textwidth}}
    \hline
    & \textbf{Approach} & \textbf{Precision}&\textbf{Recall} & \textbf{F1} & \textbf{$\Delta$ F1} \\ \hline
    \multirow{4}{*}{\rotatebox[origin=c]{90}{OSPR}}
           & ST SATD** & 0.977 & 0.324 & 0.486 & 
           \\ 
           & ST vuln.** & 0.979 & 0.958 & 0.968 & 
           \\ \cline{2-6}
           & ST SATD & 0.991 & 0.515 & 0.678 & \textcolor{ForestGreen}{$\blacktriangle$} ~0.192 
           \\ 
           & ST vuln. & 0.976 & 0.960 & 0.968 & ~~~~0.000 
           \\ 
           \hline 
           \multirow{4}{*}{\rotatebox[origin=c]{90}{Devign}}
           & ST SATD** & 0.886 & 0.846 & 0.866 & 
           \\ 
           & ST vuln.** & 0.594 & 0.589 & 0.592 & 
           \\ \cline{2-6}
           & ST SATD & 0.977 & 0.973 & 0.976  & \textcolor{ForestGreen}{$\blacktriangle$} ~0.110 
           \\
           & ST vuln. & 0.584 & 0.626 & 0.604  & \textcolor{ForestGreen}{$\blacktriangle$} ~0.012 
           \\ 
           \hline
           \multirow{4}{*}{\rotatebox[origin=c]{90}{Big-Vul}} 
           & ST SATD** & 0.931 & 0.795 & 0.858 & 
           \\ 
           & ST vuln.** & 0.934 & 0.900 & 0.912 & 
           \\ \cline{2-6}
           & ST SATD & 0.980 & 0.941 & 0.960 & \textcolor{ForestGreen}{$\blacktriangle$} ~0.102
           \\ 
           & ST vuln. & 0.944 & 0.880 & 0.911 & \textcolor{red}{$\blacktriangledown$} ~0.001 
           \\ 
    \hline  
    \multicolumn{6}{l}{**Input with internal comments left inside the function (\textit{in}).}
    \end{tabular}
\end{table}

\begin{figure}[h!t]
    \centering
    \caption{F1 for single-task SATD and vulnerability comment in and out.}
    \label{fig:ST_comments_in_vs_out}
    \includegraphics[width=\linewidth]{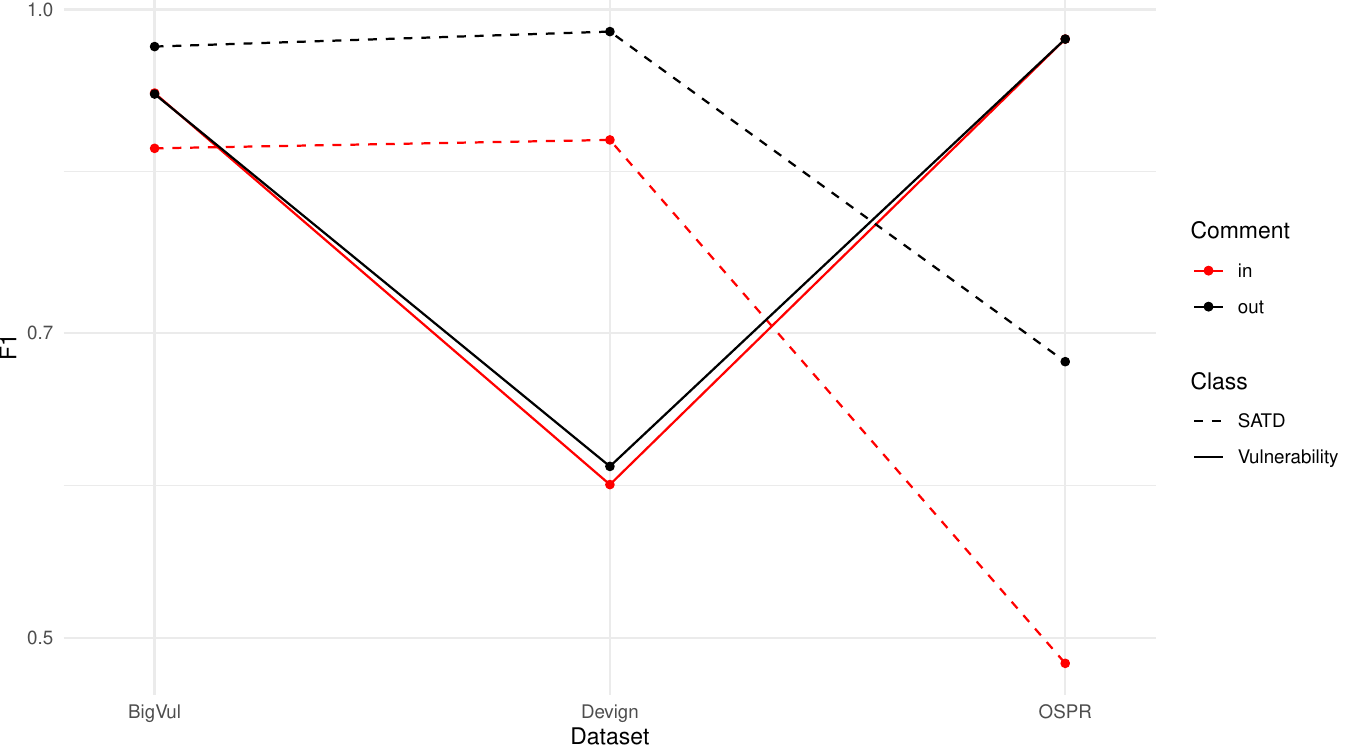}
\end{figure}

We hypothesise that removing the comments from the body of the function and adding them to the leading comment would help VulSATD distinguish comments from code and perform a more accurate classification. 
To this end, we applied VulSATD as single-task learner to our dataset and compared the performance in the two cases. For SATD classification, we find that removing internal comments (out) significantly improves VulSATD's performance compared to leaving them in (in) (Table~\ref{tab:ST_comments_in_vs_out}), even when evaluated on individual datasets (Fig.~\ref{fig:ST_comments_in_vs_out}). A similar but less pronounced trend was noted for vulnerability classification. This finding suggests that removing internal comments does not negatively impact the classification of a function's vulnerability but instead helps to better distinguish SATD comments. Therefore, in the remainder of this paper, internal comments are removed and aggregated into the leading comments. 

\subsubsection{Tokenization of the input}\label{sec:overview}
We used Byte Pair Encoding (BPE)~\cite{SennrichEtAl2016} to tokenize the input. BPE splits words into sequences of characters and identifies the most frequent symbol pair that should be merged into a new symbol. In this way, it is able to split rare words into meaningful sub-words while keeping the common words intact~\cite{Liu2019}.
The use of BPE sub-word tokenization helps to reduce the vocabulary size when tokenizing various code elements because it will split rare names (e.g., function/variables names) into multiple sub-components instead of adding the full name into the dictionary directly.
We tokenize both comments and functions. 
\subsection{CodeBERT fine-tuning}\label{sec:finetuning}
We fine-tuned CodeBERT to better capture lexical and logical semantics for the C programming language and generate a meaningful vector representation for our problem.
As suggested in Sun \etal~\cite{SunEtAL2019}, we fine-tune CodeBERT by 1) choosing a strategy to cut long input, 2) selecting the last layer of CodeBERT for the classification task, and  3) accurately selecting the hyper-parameters. 

To cut long input text, we have adopted the \textit{head-only} strategy as defined in \cite{SunEtAL2019}. The strategy simply cuts the input pair $\{$comment and function$\}$ to the first 510 tokens (i.e., maximum capacity for CodeBERT). It starts cutting the representation sequence of tokens, which is the longest between comment and function. Once the two tokens' sequences reach the same size, it keeps on cutting tokens from one and the other sequence alternatively until the total size of the two sequences has reached the maximum capacity for CodeBERT input. Given that functions are typically longer than comments, this approach may likelier cut the bottom lines of the functions' code while keeping the MAT keywords in comments as they typically appear at the beginning of a comment (e.g., \textit{FIXME} in Listing~\ref{lst:example}). 
Different layers of CodeBERT capture different levels of semantic and syntactic information with the last layer containing more general information. We selected the last layer of CodeBERT to which connect the classification component, as it has been shown that this setting gives the best performance on code classification~\cite{SunEtAL2019}. 

To define the optimal values for hyper-parameters, we perform a sensitivity analysis on the OSPR subset of MADE-WIC as it has the largest number of positive instances and trained the model with several combinations of the hyper-parameters' values, learning rate, number of epochs, dropout rate, batch size, and L2 lambda. It is interesting to note that we reached the same values used in LineVul for vulnerability detection~\cite{Fu2022}.

\begin{table}[t]
\renewcommand{\arraystretch}{1.2}
\centering
\caption{Evaluated and selected hyper-parameters as recommended by Sun~\etal~\cite{SunEtAL2019}.}
\label{tab:hyperparameters}
\begin{tabularx}{\columnwidth}{lXc}
\hline
\textbf{Hyperparameter }& \textbf{Evaluated Values} & \textbf{Selected value}\\ \hline
Learning rate & $2*10^{-5}$, $5*10^{-5}$, and $1*10^{-4}$& $2*10^{-5}$\\
Number of epochs & Up to 30 & 10\\
Dropout rate & 0, 0.1, 0.2, and 0.5 & 0.1\\
Batch size & 16 and 32 &16\\
L2 lambda & 0.0, 0.1 and 0.2 &0\\ \hline
\end{tabularx}
\end{table}

Table~\ref{tab:hyperparameters} lists all the values evaluated and the chosen one (last column). 
We split the dataset into training, validation, and testing, according to the commonly used proportion 80$\%$-10$\%$-10$\%$, \cite{SteenhoekEtAl2023, Guo2021,Fu2022}. Then, we trained for all the possible combinations and selected the parameters with the highest F1 for the validation part. 
\paragraph*{Data imbalance}\label{sec:dataImbalance}
 As in literature~(\cite{Potdar2014,Bavota2016, Ren2019TOSEM}), SATD and vulnerable functions represents only a small fraction of the functions of our dataset, Table~\ref{tab:statistics}. To account for such imbalance, we implemented a weighted loss function, the weight being the inverse of the frequency of the class. 
 
\subsection{Performance measures}\label{sec:performanceMeasuers}
We split each of the datasets in the ratio 80$\%$-10$\%$-10$\%$ and trained, validated and tested on the respective subsets. 
We applied VulSATD to classify pairs as SATD and vulnerable and SATD only or vulnerable only. To this aim, we explored both the single-task (Fig.~\ref{fig:singleTaskArch}) and multitask architectures of VulSATD (Fig.~\ref{fig:multiTasksArch}).
At step \circled{3}, we fine-tune the model.
Finally, we analyse the results of all the experiments in terms of precision, recall and their harmonic mean F1:

\begin{equation}
    \text{Precision} = \frac{TP}{TP+FP}
\end{equation}
\\
\begin{equation}
    \text{Recall} = \frac{TP}{TP+FN}
\end{equation}
\\
\begin{equation}
    \text{F1} = 2*\frac{\text{Precision}*\text{Recall}}{\text{Precision}+\text{Recall}}
\end{equation}\\

\subsection{Implementation details and replication package}
\label{sec:replicationPackage}

We implemented the proposed approach using TensorFlow 2.0 and Keras in Python 3. We ran the tasks in a cluster consisting of two Nvidia A100 GPUs with 192 GB of RAM, in a server with the processor Xeon 4208 with 16 cores per node, i.e., 32 cores in total.
The maximum time of execution for the largest dataset (OSPR) was around 104ks ($\sim$ 29 hours) to train and validate and 513s to test. 

The code and the script for mutation of the dataset used for this work are publicly available in our replication package~\cite{RussoEtAl2024Replication}.

\section{Experimental design}
\label{sec:evaluation}

To fulfil our research goal, we performed a series of experiments with VulSATD to answer the following research questions:\\

\newcommand{\rqone}{\noindent\textbf{RQ1: Can multi-task  improve  single-task learning of low-quality code?}\par}
\rqone

Our initial hypothesis is that both TD and vulnerability are forms of suboptimal code that reduce the overall quality and maintainability of a software system. TD is often a deliberate trade-off, where developers knowingly prioritize speed over quality (e.g., skipping refactoring). Weak code, on the other hand, is usually unintentional and stems from poor coding practices, lack of knowledge, or oversight. While TD requires more effort to fix if addressed at later stages, weak code must be fixed as soon as possible, although it may not be discovered until much later (e.g., zero-day vulnerabilities).
Russo \etal ~\cite{Russo2022} found that technical debt and weak code co-occur in 55\% of C files. They and Ferreyra \etal~\cite{FerreyraEtAl2024} also found that security is a transversal concern in technical debt. Thus, we hypothesize that:

\textit{H1: The two forms of low-quality code may share common information, and one can serve as important clues for the other}, which motivates us to employ multi-task learning for the prediction.  

Thus, we investigate this hypothesis by applying VulSATD with the single-task and multi-task architecture, Figs. \ref{fig:singleTaskArch} and \ref{fig:multiTasksArch}.
 For each configuration, we compute Precision, Recall and F1 as defined in Section \ref{sec:performanceMeasuers}.  We finally compute the delta difference of F1 between the two classifications across the datasets. 

\newcommand{\rqtwo}{\noindent\textbf{RQ2: Is the imbalance nature of the annotated dataset affecting the performance of the multi-task learning?}\par}\rqtwo

Class imbalance, i.e., when one category on a classification setting is represented by a minority of instances, has been an issue for machine learning, even with deep learning approaches~\cite{Ghosh2024,Krawczyk2016}.  With this research question, we investigate whether using a weighted loss — a common strategy to address class imbalance — can enhance the performance of VulSATD.
Thus, we hypothesize that:

\textit{H2: Data imbalance for SATD and vulnerability classification affects the performance of VulSATD in any architecture version}, which motivates the comparison of the VulSATD models with regular and weighted loss functions.  

Thus, we investigate this hypothesis by applying VulSATD with the single-task and multi-task architecture, Figs. \ref{fig:singleTaskArch} and \ref{fig:multiTasksArch} and weighted loss.
 For each configuration, we compute Precision, Recall and F1 as defined in Section \ref{sec:performanceMeasuers}. We finally compute the delta difference of F1 between the two classifications across the datasets and the delta difference of F1 between the weighted and regular loss. 

\section{Results}
\label{sec:results}

In this section, we answer the research question and discuss the hypothesis we made.

\rqone

To answer this question, we compare the performance of VulSATD in its two architectures and each of the datasets of MADE-WIC as described in Tables ~\ref{tab:datasets} and ~\ref{tab:statistics}.
We split each dataset into training/validation/test subsets with 80$\%$-10$\%$-10$\%$ proportion.
Table \ref{tab:fixed_loss_MT_vs_ST} reports F1, Precision, and Recall of the test set for all datasets, the two architectures (MT multi-task and ST single task), and the classification of vulnerability and SATD. The last column of the table indicates the increase or decrease of F1 for the multi-task vs. the single-task architecture. For instance, for the dataset, OSPR, F1 increases by 0.033 with the multi-task architecture for the SATD classification (MT SATD vs. ST SATD) and decreases by 0.001 for the classification of vulnerable functions (MT vuln. vs. ST vuln.). Overall, we can see that the increase or decrease in F1 is minimal even in the case in which room for improvement is possible (i.e., Devign classification of vulnerable functions). There is no specific trend across the projects as Big-Vul has opposite trend of $\Delta$ F1 with respect the other two datasets both for SATD and vulnerability classification. 
\begin{table}[b]
\renewcommand{\arraystretch}{1.2}
\caption{Precision, Recall, and F1 for single-task and multi-task classification of functions for SATD and vulnerability.  $\Delta$ F1 compares multi-task against the single-task learning.}
    \label{tab:fixed_loss_MT_vs_ST}
    \centering
    \begin{tabular}{clC{0.06\textwidth}C{0.06\textwidth}C{0.06\textwidth}C{0.06\textwidth}}
    \hline
    & \textbf{Approach} & \textbf{Precision}&\textbf{Recall} & \textbf{F1} & \textbf{$\Delta$ F1} \\ \hline
    \multirow{4}{*}{\rotatebox[origin=c]{90}{OSPR}} & MT SATD & 0.924 & 0.578 & 0.711 &  \textcolor{ForestGreen}{$\blacktriangle$} ~0.033\\ 
           & MT vuln. & 0.975 & 0.958 & 0.967 & \textcolor{red}{$\blacktriangledown$} -0.001 
           \\
           \cline{2-6} 
           & ST SATD & 0.991 & 0.515 & 0.678 & 
           \\ 
           & ST vuln. & 0.976 & 0.960 & 0.968 & 
           \\
    \hline 
    \multirow{4}{*}{\rotatebox[origin=c]{90}{Devign}} & MT SATD & 0.989 & 0.985 & 0.987 & \textcolor{ForestGreen}{$\blacktriangle$} ~0.011
      \\  
           & MT vuln. & 0.601 & 0.567 & 0.584 & \textcolor{red}{$\blacktriangledown$} -0.020\\
           \cline{2-6}
           & ST SATD & 0.977 & 0.973 & 0.976 & 
           \\
           & ST vuln. & 0.584 & 0.626 & 0.604 & 
           \\ 
    \hline
    \multirow{4}{*}{\rotatebox[origin=c]{90}{Big-Vul}} & MT SATD  & 0.970 & 0.936 & 0.953 & \textcolor{red}{$\blacktriangledown$} -0.007
    \\
           & MT vuln. & 0.948 & 0.880 & 0.913 & \textcolor{ForestGreen}{$\blacktriangle$} ~0.002 \\
           \cline{2-6}
           & ST SATD & 0.980 & 0.941 & 0.960 & 
           \\
           & ST vuln. & 0.944 & 0.880 & 0.911 & 
           \\
    \hline  
    \end{tabular}
\end{table}

The results indicate that sharing the CodeBERT layer between tasks (e.g., SATD and vulnerability classification) does not impair the model's classification performance.  This setup eliminates the need to run separate models, making it a resource-efficient solution. We indeed compared the speed of computation of single and multi-tasks classification on the OSPR dataset. \textit{We found  that running multi-task learning is twice as fast both in  training and  test  the data.} This suggests that in environments with limited computational resources, a  CodeBERT instance with multi-task architecture could be preferable.

It is also worth noting our results are consistent with existing literature. For instance, the single-task architecture we used as well as the results for single-task vulnerability classification we obtain in Table~\ref{tab:weighted_vs_normal_loss} mirrors the ones of the LineVul model~\cite{FuTantithamthavorn2022}. 

\summarybox{The hypothesis \textit{H1: The two forms of low-quality code may share common information, and one can serve as important clues for the other} cannot be confirmed, as no difference in performance between the multi-task and single-task classifications is observed. The results suggest that the information embedded in SATD comments does not enhance VulSATD's ability to classify vulnerabilities in the functions of any of the MADE-WIC datasets, nor does the information from vulnerabilities improve the classification of SATD.}

\rqtwo

We repeated the same process for the previous RQ but using weighted loss during the training of the models. The results are presented in Table~\ref{tab:weighted_vs_normal_loss}. The second last column reports the increase or decrease of F1 for the multi-task vs. the single-task architecture.  The differences observed are again small (the greatest is F1=0.033 for multi-task SATD classification) and not consistent over the datasets for both SATD and vulnerability classifications (see little arrows). The last column in Table~\ref{tab:weighted_vs_normal_loss} shows $\Delta'$ F1, the difference between the use of the  weighted loss with the regular loss function. Again the difference is very small (the greatest is F1=0.046 for the SATD classification in Big-Vul) and the trend changes over datasets and classification task. 

\begin{table}[ht]
\renewcommand{\arraystretch}{1.2}
    \caption{Precision, Recall, and F1 for single-task SATD and vulnerability, and multi-tasks SATD and vulnerability, with weighted loss. $\Delta$ F1 compares multi-task against single-task classification. $\Delta'$ F1 compares  F1 of Table~\ref{tab:fixed_loss_MT_vs_ST} : weighted vs. regular loss.}
    \label{tab:weighted_vs_normal_loss}
    \centering
    \begin{tabular}{clC{0.04\textwidth}C{0.04\textwidth}C{0.04\textwidth}C{0.06\textwidth}C{0.06\textwidth}}
    \hline
    & \textbf{Approach} & \textbf{Precision}&\textbf{Recall} & \textbf{F1} & \textbf{$\Delta$ F1}&\textbf{$\Delta'$ F1}\\ \hline
    \multirow{4}{*}{\rotatebox[origin=c]{90}{OSPR}} & MT SATD  & 0.975 & 0.535 & 0.690 & \textcolor{ForestGreen}{$\blacktriangle$} ~0.033&\textcolor{red}{$\blacktriangledown$}~-0.021\\ 
           & MT vuln. & 0.971 & 0.966 & 0.969 &\textcolor{red}{$\blacktriangledown$}~-0.001& \textcolor{ForestGreen}{$\blacktriangle$} ~0.002\\ 
           \cline{2-7} 
           & ST SATD & 0.911 & 0.585 & 0.713 & &\textcolor{ForestGreen}{$\blacktriangle$} ~0.035\\ 
           & ST vuln. & 0.974 & 0.960 & 0.967 & &\textcolor{red}{$\blacktriangledown$}~-0.001\\ 
    \hline 
    \multirow{4}{*}{\rotatebox[origin=c]{90}{Devign}} 
           & MT SATD & 0.988 & 0.944 & 0.966 & \textcolor{ForestGreen}{$\blacktriangle$} ~0.011&\textcolor{red}{$\blacktriangledown$} -0.021\\ 
           & MT vuln. & 0.598 & 0.556 & 0.576 & \textcolor{red}{$\blacktriangledown$}~ -0.02&\textcolor{red}{$\blacktriangledown$} -0.008\\
           \cline{2-7}
           & ST SATD & 0.985 & 0.974 & 0.979 & &\textcolor{ForestGreen}{$\blacktriangle$} ~0.003\\
           & ST vuln. & 0.583 & 0.582 & 0.583 & &\textcolor{red}{$\blacktriangledown$} -0.021\\ 
    \hline 
    \multirow{4}{*}{\rotatebox[origin=c]{90}{Big-Vul}} & MT SATD & 0.965 & 0.941 & 0.953 &\textcolor{red}{$\blacktriangledown$}~-0.007 &\textcolor{red}{$\blacktriangledown$} -0.046\\ 
           & MT vuln. & 0.928 & 0.888 & 0.908 & \textcolor{red}{$\blacktriangledown$}~0.002 &\textcolor{red}{$\blacktriangledown$} -0.005\\
           \cline{2-7}
           & ST SATD & 0.946 & 0.941 & 0.944 & &\textcolor{red}{$\blacktriangledown$} -0.016 \\ 
           & ST vuln. & 0.941 & 0.895 & 0.918 &  &\textcolor{ForestGreen}{$\blacktriangle$} ~0.007\\ 
    \hline  
    \end{tabular}
\end{table}

\summarybox{The hypothesis \textit{H1: Data imbalance for SATD and vulnerability classification affects the
performance of VulSATD in any architecture version.} cannot be confirmed, as no difference in performance between the multi-task and single-task classifications is observed with or without balancing the learning. The results suggest again that the information embedded in SATD comments does not enhance VulSATD’s ability to classify vulnerabilities in the functions of any of the MADE-WIC datasets, nor does the information from vulnerabilities improve the classification of SATD.}

\section{Discussion}
\label{sec:discussion}
Based on our results, the multi-task architecture does not improve the classification of TD or vulnerable functions. 
This may be due to different reasons (see Section~\ref{sec:threats}).  
One of them can be the general SATD annotation. 
Listing \ref{lst:example_non_vul_SATD} and \ref{lst:example_vul_SATD} illustrate two functions that are annotated as TD by the SATD comment (Listing ~\ref{lst:example_non_vul_SATD}, line 17 and Listing ~\ref{lst:example_vul_SATD}, line 9). 
According to the annotation of MADE-WIC/Devign dataset, the first is non-vulnerable, whereas the second is vulnerable. However, the first uses \texttt{av\_malloc} function (Listing ~\ref{lst:example_non_vul_SATD}, line 18) that might be exploited for buffer overflow. The second uses \texttt{memset} function (Listing ~\ref{lst:example_vul_SATD}, line 12) that again might be exploited for buffer overflow.
This subtle difference could not be inferred from the annotation neither adjusted with the shared information from the SATD comment. 
To understand whether the richer information on the type of SATD can be considered in future work, we have analysed the distribution of SATD types over vulnerable and non-vulnerable functions. To this aim, we have extracted\footnote{100 vulnerable functions are randomly extracted and the other 100 are taken from their fixed commit as per the approach of Devign} 200 SATD-annotated functions of the Devign dataset \cite{ZhouEtAl2019}, of which half was vulnerable and the other half not. Then, we leveraged the existing taxonomy of SATD comments~\cite{Bavota2016,Alves2014} and applied it to the sample. 
From the SATD taxonomy~\cite{Bavota2016}, we found instances for the following categories: \textit{design debt, requirement debt, code debt, test debt, and defect debt}. 
\begin{lstlisting}[
    basicstyle=\footnotesize,
    escapechar=!, 
    language=C, 
keywordstyle=\color{blue}, 
    breaklines=true, 
    numbers=left,
    label={lst:example_non_vul_SATD},
    caption={Example of a non-vulnerable function containing a SATD comment, extracted from the Devign dataset.},
    xleftmargin=2em,
    framexleftmargin=1.5em,
    tabsize=1,
    numbersep=3pt,
    columns=fullflexible,
    float=b
]
/**
 * av_realloc semantics (same as glibc): if ptr is NULL and size > 0,
 * identical to malloc(size). If size is zero, it is identical to
 * free(ptr) and NULL is returned.
 */
void *av_realloc(void *ptr, unsigned int size)
{
#ifdef MEMALIGN_HACK
    int diff;
#endif

    /* let's disallow possible ambiguous cases */
    if(size > INT_MAX)
        return NULL;

#ifdef MEMALIGN_HACK
    //FIXME this isn't aligned correctly, though it probably isn't needed
    if(!!ptr) return av_malloc(size);
    diff= ((char*)ptr)[-1];
    return realloc(ptr - diff, size + diff) + diff;
#else
    return realloc(ptr, size);
#endif
}
\end{lstlisting}
\begin{lstlisting}[
    basicstyle=\footnotesize,
    escapechar=!, 
    language=C, 
keywordstyle=\color{blue}, 
    breaklines=true, 
    numbers=left,
    label={lst:example_vul_SATD},
    caption={Example of a vulnerable function containing a SATD comment, extracted from the Devign dataset.},
    xleftmargin=2em,
    framexleftmargin=1.5em,
    tabsize=1,
    numbersep=3pt,
    columns=fullflexible,
    float=ht
]
static void vscsi_process_login(VSCSIState *s, vscsi_req *req)
{
    union viosrp_iu *iu = &req->iu;
    struct srp_login_rsp *rsp = &iu->srp.login_rsp;
    uint64_t tag = iu->srp.rsp.tag;

    trace_spapr_vscsi__process_login();

    /* TODO handle case that requested size is wrong and
     * buffer format is wrong
     */
    memset(iu, 0, sizeof(struct srp_login_rsp));
    rsp->opcode = SRP_LOGIN_RSP;
    /* Don't advertise quite as many request as we support to
     * keep room for management stuff etc...
     */
    rsp->req_lim_delta = cpu_to_be32(VSCSI_REQ_LIMIT-2);
    rsp->tag = tag;
    rsp->max_it_iu_len = cpu_to_be32(sizeof(union srp_iu));
    rsp->max_ti_iu_len = cpu_to_be32(sizeof(union srp_iu));
    /* direct and indirect */
    rsp->buf_fmt = cpu_to_be16(SRP_BUF_FORMAT_DIRECT | SRP_BUF_FORMAT_INDIRECT);

    vscsi_send_iu(s, req, sizeof(*rsp), VIOSRP_SRP_FORMAT);
}


\end{lstlisting}
Table \ref{tab:Characterization} illustrates the different categories for vulnerable and non-vulnerable functions in the sample. Design and Code debt are the most frequent TD functions, followed by Requirement debt. 
The number of vulnerable functions is greater in Code debt, whereas it is smaller for Design debt. The result on Code debt is in line with the work of Bavota and Russo \cite{Bavota2016}
whereas the number of instances of Requirement debt is greater in our sample.  
Of course, the distribution in the Table can be specific to the sample we randomly choose, but it indicates that further work in this direction is needed. 
\begin{table}[!ht]
    \renewcommand{\arraystretch}{1.2}
    \caption{SATD types distribution over a set of 200 SATD annotated functions balanced over vulnerability.}
    \label{tab:Characterization}
    \centering
    \begin{tabular}{lccc}
        \hline
        \textbf{Characteristic} & \textbf{\# vul.} & \textbf{\# no vul.} & \textbf{Total}\\ 
        \hline
        \textit{Design debt} & 29 & 38 &67\\
        \textit{Requirement debt} & 16 & 21 & 37\\
        \textit{Code debt} & 43 & 27 & 70\\
        \textit{Test debt} & 0 & 2 & 2\\ 
        \textit{Defect Debt} & 12 & 12& 24\\ 
        \hline
        \textit{Total}&100&100 & 200\\
        \hline
    \end{tabular}
\end{table}

The information contained in SATD comments may not always be highly informative, as their inclusion is a voluntary action by developers. We observed instances where the semantic usage of the four MAT patterns is sometimes misapplied or lacks descriptiveness,  as illustrated in Listing \ref{lst:example_misusage}. The listing shows a comment containing only the MAT pattern ``FIXME", which is also used incorrectly: all authors agreed that ``TODO" or ``XXX" would be a more fitting descriptor for a stub rather than ``FIXME".

\begin{lstlisting}[
    basicstyle=\footnotesize,
    escapechar=!, 
    language=C, 
keywordstyle=\color{blue}, 
    breaklines=true, 
    numbers=left,
    label={lst:example_misusage},
    caption={Example of misusage for one of the MAT patterns, extracted from the Devign dataset.},
    xleftmargin=2em,
    framexleftmargin=1.5em,
    tabsize=1,
    numbersep=3pt,
    columns=fullflexible,
    float=bh
]
void qpci_iounmap(QPCIDevice *dev, void *data)

{

    /* FIXME */

}
\end{lstlisting}
\section{Related Work}
\label{sec:relatedWork}

In this section, we discuss related work regarding vulnerability detection (Section \ref{ssec:vulnerabilities}) and SATD detection (Section \ref{ssec:SATD}). 
\begin{table*}[htb]
    \centering
    \caption{Performance of the SOTA methods and their original datasets in comparison with the best results of VulSATD on MADE-WIC datasets. The value for each category (SATD and Vulnerability) is highlighted.}
    \label{tab:performanceComparison}
    \renewcommand{\arraystretch}{1.2}
    \begin{tabular}{p{0.10\textwidth}p{0.45\textwidth}p{0.2\textwidth}p{0.05\textwidth}}
       \hline \textbf{Method Name}&\textbf{Method} & \textbf{Dataset} &  \textbf{F1}\\
        \hline
        \multicolumn{4}{c}{\textbf{SATD}}\\
        \hline
       -&Maximum Entropy classifier~\cite{Maldonado2017}& 10 Java projects~\cite{Maldonado2017}&0.62\\
       -&Naive Bayes classifier~\cite{Huang2018}&8 open-source projects&0.74\\
-&Convolutional Neural Network ~\cite{Ren2019TOSEM}&10 Java projects \cite{Maldonado2017}&0.77\\
     -&BERT  ~\cite{Aiken2023}&20 Java projects ~\cite{Guo2021}&0.87\\
       HATD& Embedding from Language Models with Hybrid attention matrix &10 Java projects \cite{Maldonado2017}&0.83\\
       Jitterbug &Hybrid: pattern based and machine learning~\cite{Yu2022TSE}&10 Java projects~\cite{Maldonado2017}&0.43\\
       VulSATD&CodeBERT with multi-task classification and weighted loss function&MADE-WIC/Devign&0.96\\
       VulSATD&CodeBERT with single-task classification and weighted loss function&MADE-WIC/Devign&\textbf{0.98}\\
       \hline
       \multicolumn{4}{c}{\textbf{Vulnerability}}\\
       \hline
       IVDetect&Graph Convolutional Neural Network ~\cite{Li2021FSE} &ReVeal (subset of Devign) \cite{Chakraborty2022}&0.45\\
      &&Fan~\cite{Fan2020}&0.35\\
      && FFMpeg \& Qemu~\cite{ZhouEtAl2019}&0.65\\       
      REVEAL&Convolutional Neural Network+RF~\cite{Chakraborty2022}&REVEAL&0.41\\
      &&FFMpeg+Qemu&0.64\\
       
      LineVul& CodeBERT ~\cite{Feng2020} ~\cite{Fu2022} & Big-Vul \cite{Fan2020} &0.91 \\
      Devign&Gated Graph Convolutional Neural Network ~\cite{ZhouEtAl2019} &Devign (four C projects) &  0.85\\
      VulSATD&CodeBERT with multi-task classification and weighted loss function&MADE-WIC/OSPR&\textbf{0.97}\\
      VulSATD&CodeBERT with single-task classification and regular loss function&MADE-WIC/OSPR&\textbf{0.97}\\
      \hline
    \end{tabular}
\end{table*}

\subsection{Vulnerability Detection}
\label{ssec:vulnerabilities}

Employing machine and deep learning approaches to detect vulnerabilities has been vastly explored in the literature. Zhou \etal~\cite{ZhouEtAl2019} proposed Devign, an approach based on graph neural networks, to detect vulnerable functions. To evaluate their approach, the authors manually labelled a dataset extracted from four large open-source projects in C: Linux Kernel, QEMU, FFmpeg, and Wireshark. Li \etal~\cite{Li2021FSE} proposed IVDetect, a vulnerability detection model based on graph convolutional neural networks, aiming to detect vulnerable functions but also to tell which statements are responsible for the vulnerability, increasing the explainability of the results. Their results outperformed previous approaches based on deep learning by employing program dependency graphs (PDGs). They evaluated their approach in three datasets: ReVeal~\cite{Chakraborty2022}, the fraction of Devign~\cite{ZhouEtAl2019} containing the projects QEMU and FFmpeg, and Big-Vul~\cite{Fan2020}. 

Fu \etal~\cite{Fu2022} proposed LineVul, a line-level vulnerability prediction approach based on the BERT architecture. They built the model based on the pre-trained CodeBERT~\cite{Feng2020} and compared it with IVDetect, obtaining better results. It is interesting to note that LineVul does not need different code representations as IVDetect or Devign, only relying on CodeBERT tokenization and representation.

Chakraborty \etal~\cite{Chakraborty2022} investigated how SOTA deep-learning-based techniques behaved in a real-world scenario. To do so, they curated a dataset based on Chromium and the Linux Debian Kernel, based on code patches labelled as security.

\subsection{SATD Detection}
\label{ssec:SATD}

Potdar and Shihab~\cite{Potdar2014} proposed the term Self-Admitted Technical Debt to identify source code comments pointing to instances of technical debt. Since then, the term was extended to other natural language artefacts associated with software development, such as issues~\cite{Xavier2020,Li2022ese}. In this work, we stitch with the initial definition and focus on source code comments tagging code. 

Research on SATD can be grouped in three categories~\cite{Sierra2019}: detection, comprehension, and repayment. Detection approaches aimed to determine if source code comments were SATD or not, and they can be classified into pattern-based or machine learning-based approaches.
Pattern-based approaches have the advantages of easy implementation and replicability~\cite{Sierra2019,Guo2021}, with the drawback of increased false positives~\cite{Bavota2016}. In the paper presenting SATD~\cite{Potdar2014}, Potdar and Shihab employed 62 patterns to identify SATD in 2.4\% to 31\% of the files in four large open-source software projects: Eclipse, Chromium OS, Apache HTTP Server, and ArgoUML. In a larger study considering 159 projects, Bavota and Russo \cite{Bavota2016} estimated that this approach led to around 25\% of false positives. 
Machine learning-based approaches were suggested to tackle this issue with the drawback of the need for a labelled dataset.
Maldonado \etal~\cite{Maldonado2017} employed a natural language processing (NLP) maximum entropy classifier. The authors also built a dataset by extracting and manually labelling comments from ten Java open-source projects. By performing a cross-project evaluation, training in nine projects and testing on the other for each project, they reached an average F1 of 0.62. Still relying on NLP techniques, Huang \etal~\cite{Huang2018} proposed a naive Bayes classifier for the SATD detection problem. The authors evaluated their approach in 8 open-source projects, obtaining an average F1 of 0.74. Ren \etal~\cite{Ren2019TOSEM} proposed an approach based on convolutional neural networks (CNN). They evaluated it with the dataset of Maldonado \etal~\cite{Maldonado2017}, reaching an average F1 of 0.77. They also run the naive Bayes classifier on this dataset, obtaining an average F1 of 0.7. 

A major issue with the machine learning-based approaches has been the replication of the results. Guo \etal~\cite{Guo2021} investigated this problem by trying to replicate the three above-mentioned approaches: maximum entropy, naive Bayes, and CNN, using Maldonado and Shihab's dataset. Regarding the CNN approach, they were not able to replicate the results obtained by Ren \etal~\cite{Ren2019TOSEM}. They also proposed a new pattern-based approach, called Matches task Annotation Tags (MAT), based on four task annotation tags, i.e., ``TODO'', ``FIXME'', ``XXX'', and ``HACK'', obtaining similar results to the CNN approach. Another contribution of the study was the extension of the dataset by extracting and manually labelling the comments of ten other open-source projects in Java.  

Following the emergence of attention-based mechanisms for machine learning~\cite{Vaswani2002}, Wang \etal~\cite{Wang2020} proposed HATD (Hybrid Attention-based method for self-admitted technical debt) using ELMo (Embedding from Language Models). By evaluating Maldonado and Shihab's dataset, they reached an average F1 of 0.83. In a recent study, Aiken \etal~\cite{Aiken2023} fine-tuned BERT to the SATD detection task. 
They reached an average F1 of 0.86 on a cross-project evaluation in Maldonado and Shihab's dataset and 0.87 in Guo et al.'s extension. Besides supervised learning approaches, researchers have explored the possibility of using semi-supervised or active learning approaches. Yu \etal~\cite{Yu2022TSE} proposed Jitterbug by first detecting ``easy'' SATDs, using words similar to MAT, then using machine learning approaches to help humans decide the final classification. Similarly, Tu \etal~\cite{Tu2022} proposed DebtFree, a two-phase approach, where the first step is an unsupervised approach based on CLA (Clustering and Labeling)~\cite{Nam2016} and the second step is active learning on more difficult labels. They reached similar results to the CNN approach but with a smaller labelling effort and use Recall and Cost to compare their solution with literature.

In summary, several approaches have been proposed to identify SATD in an automatic way. Machine learning-based approaches generally have better results with the expense of labelled datasets. In this regard, most of the studies relied on the dataset provided by Maldonado \etal~\cite{Maldonado2017} consisting of source comments of ten open-source Java projects. Therefore, our study presented some innovations compared to the literature. First, although Aiken \etal~\cite{Aiken2023} employed BERT, to the best of our knowledge, no approach was based on CodeBERT that has been trained with code. Second, no supervised learning approach employed the information regarding to support SATD detection. DebtFree uses a proxy for code complexity, but it is a semi-supervised learning. 

Finally, in Table~\ref{tab:performanceComparison}, we have reported the best values obtained with VulSATD both for SATD and vulnerability classification. From the table, we can see that:
\summarybox{
VulSATD is outperforming existing works on SATD or vulnerability detection using both multi-task and single-task architecture. }

\section{Threats to validity}\label{sec:threats}

\textit{Threats to construct validity} are mainly related to the construction of the datasets and the input preparation. 
Firstly, we rely on the annotations of an existing dataset, MADE-WIC. Although the dataset is recognized by the research community, the classification may suffer for its specific annotations. For vulnerability classification, this can be seen in Listings ~\ref{lst:example_non_vul_SATD} and \ref{lst:example_vul_SATD}). To mitigate such an aspect, we have run our analysis on the different datasets of MADE-WIC. When we use SATD to classify functions that contain TD, we are also forgetting functions that have TD, but developers have not annotated it in their comments. Future work will dig into different ways to detect TD, such as by detecting code smells. 
Secondly, the datasets turn out to be imbalanced. To mitigate this aspect, we have repeated the learning with a weighted loss function. This resulted in simpler than balancing the sample for the four classes and two architectures. Our results are already outperforming existing literature, leaving, in some cases, little room for improvement.  
Thirdly, the input to CodeBERT has been cut according to the \textit{head} strategy for which the input tail is cut until 510 tokens. The cut may have removed important information from the input. For instance, it may have removed code lines related to a vulnerability pattern in a function. Future work will explore the \textit{head} and \textit{tail} strategy that seems to be winning for general training of the CodeBERT model.

\textit{Threats to internal validity} concern factors internal to our studies that could have influenced our results.
As the definitions of vulnerability and technical debt are themselves not unique, the techniques used to annotate the datasets may have been inconsistent. For instance, the annotation for OSPR did not contain regular expressions to find all CWE vulnerabilities in code written in C, neither the change set of a fixing commit can assure that the line of codes that have been changed pertains to a vulnerability. For this reason, we extended our work to datasets with different annotation procedures.  

\textit{Threats to external validity}
To generalize our results, we used MADE-WIC, which is based on three different datasets, containing multiple annotations. 
These datasets include a large set of projects used in the research of vulnerability and SATD detection. Even though we know that this is not comprehensive, we believe that it is enough representative of the population we want to analyse. 

\section{Conclusions}
\label{sec:conclusions}

Our general objective is to provide decision-making tools that make developers aware of issues such as vulnerability and technical debt. In this work, we have discussed whether the information on one aspect influences the detection of the other aspect of low-quality code. Based on a hypothesis that SATD and vulnerabilities are both concerning ugly code that works, we investigated if a multi-task approach, leveraging the information shared between these concerns, could improve their automatic detection. 
To this aim, we have implemented VulSATD, a classifier that simultaneously detects SATD and vulnerabilities in functions. The core of the machine learner is based on the SOTA tokenizer BPE and CodeBERT, a pre-trained transformers-architecture model. 
VulSATD exploits the information carried by both comments and function code thanks to the bimodal feature of CodeBERT. 
We have designed two architectures for VulSATD, a multi-task and a single-task one. The multi-task instance classifies SATD or vulnerable functions through the shared knowledge from the comments and the function's code, the single task one classifies functions separately for SATD or vulnerability.
The results show that sharing information does not enhance VulSATD's performance. However, running multiple tasks simultaneously is twice as fast as executing a single task. Therefore, when resources are limited, a multi-tasking approach is the better option.
Even though we did not check the results for other models, since CodeBERT led to good results for both single-tasks, the fact that multi-task did not improve the results for this case provides a piece of evidence that improving the results by sharing the information of these tasks is, at least, not valid in all cases.

Finally, our tool is publicly available (Section~\ref{sec:replicationPackage}). We support open science and encourage the community to continue improving vulnerability and  SATD detection with further studies. With our work, we aim to stimulate other studies to investigate further the application of CodeBERT and multi-task classification for code-related tasks.
\\
\section*{Acknowledgment}
\label{sec:ack}
We acknowledge ISCRA for awarding this project access to the LEONARDO supercomputer, owned by the EuroHPC Joint Undertaking, hosted by CINECA (Italy). Moritz Mock is partially funded by the National Recovery and Resilience Plan (Piano Nazionale di Ripresa e Resilienza, PNRR - DM 117/2023). The work has been funded by the project  CyberSecurity Laboratory no. EFRE1039 under the 2023 EFRE/FESR program.

\bibliographystyle{IEEEtran}
\bibliography{references}   

\begin{thebibliography}{10}
\providecommand{\url}[1]{#1}
\csname url@samestyle\endcsname
\providecommand{\newblock}{\relax}
\providecommand{\bibinfo}[2]{#2}
\providecommand{\BIBentrySTDinterwordspacing}{\spaceskip=0pt\relax}
\providecommand{\BIBentryALTinterwordstretchfactor}{4}
\providecommand{\BIBentryALTinterwordspacing}{\spaceskip=\fontdimen2\font plus
\BIBentryALTinterwordstretchfactor\fontdimen3\font minus
  \fontdimen4\font\relax}
\providecommand{\BIBforeignlanguage}[2]{{%
\expandafter\ifx\csname l@#1\endcsname\relax
\typeout{** WARNING: IEEEtran.bst: No hyphenation pattern has been}%
\typeout{** loaded for the language `#1'. Using the pattern for}%
\typeout{** the default language instead.}%
\else
\language=\csname l@#1\endcsname
\fi
#2}}
\providecommand{\BIBdecl}{\relax}
\BIBdecl

\bibitem{Fowler1999}
M.~Fowler, \emph{Refactoring: Improving the Design of Existing Code}.\hskip 1em
  plus 0.5em minus 0.4em\relax Boston, MA, USA: Addison-Wesley Longman
  Publishing Co., Inc., 1999.

\bibitem{Cunningham2009}
\BIBentryALTinterwordspacing
C.~Ward, ``Ward explains debt metaphor,'' 2009. [Online]. Available:
  \url{wiki.c2.com/?WardExplainsDebtMetaphor}
\BIBentrySTDinterwordspacing

\bibitem{Izurieta2018}
C.~Izurieta, D.~Rice, K.~Kimball, and T.~Valentien, ``{A position study to
  investigate technical debt associated with security weaknesses},'' in
  \emph{Proceedings of the 2018 International Conference on Technical
  Debt}.\hskip 1em plus 0.5em minus 0.4em\relax New York, NY, USA: ACM, may
  2018, pp. 138--142.

\bibitem{Russo2022}
B.~Russo, M.~Camilli, and M.~Mock, ``Weaksatd: Detecting weak self-admitted
  technical debt,'' in \emph{19th {IEEE/ACM} International Conference on Mining
  Software Repositories, {MSR} 2022, Pittsburgh, PA, USA, May 23-24,
  2022}.\hskip 1em plus 0.5em minus 0.4em\relax {ACM}, 2022, pp. 448--453.

\bibitem{FerreyraEtAl2024}
N.~E.~D. Ferreyra, M.~Shahin, M.~Zahedi, S.~Quadri, and R.~Scandariato, ``{
  What Can Self-Admitted Technical Debt Tell Us About Security? A Mixed-Methods
  Study },'' in \emph{2024 IEEE/ACM 21st International Conference on Mining
  Software Repositories (MSR)}.\hskip 1em plus 0.5em minus 0.4em\relax Los
  Alamitos, CA, USA: IEEE Computer Society, Apr. 2024, pp. 704--715.

\bibitem{Edbert2023}
J.~A. Edbert, S.~J. Oishwee, S.~Karmakar, Z.~Codabux, and R.~Verdecchia,
  ``{Exploring Technical Debt in Security Questions on Stack Overflow},'' in
  \emph{2023 ACM/IEEE International Symposium on Empirical Software Engineering
  and Measurement (ESEM)}.\hskip 1em plus 0.5em minus 0.4em\relax IEEE, oct
  2023, pp. 1--12.

\bibitem{Zhang2022}
Y.~Zhang and Q.~Yang, ``{A Survey on Multi-Task Learning},'' \emph{IEEE
  Transactions on Knowledge and Data Engineering}, vol.~34, no.~12, pp.
  5586--5609, dec 2022.

\bibitem{Li2022}
Y.~Li, X.~Che, Y.~Huang, J.~Wang, S.~Wang, Y.~Wang, and Q.~Wang, ``{A Tale of
  Two Tasks: Automated Issue Priority Prediction with Deep Multi-task
  Learning},'' in \emph{ACM / IEEE International Symposium on Empirical
  Software Engineering and Measurement (ESEM)}.\hskip 1em plus 0.5em minus
  0.4em\relax New York, NY, USA: ACM, 2022, pp. 1--11.

\bibitem{Liu2022}
F.~Liu, G.~Li, B.~Wei, X.~Xia, Z.~Fu, and Z.~Jin, ``{A unified multi-task
  learning model for AST-level and token-level code completion},''
  \emph{Empirical Software Engineering}, vol.~27, no.~4, p.~91, jul 2022.

\bibitem{Bavota2016}
G.~Bavota and B.~Russo, ``A large-scale empirical study on self-admitted
  technical debt,'' in \emph{Proceedings of the 13th International Conference
  on Mining Software Repositories}, ser. MSR '16.\hskip 1em plus 0.5em minus
  0.4em\relax New York, NY, USA: ACM, 2016, pp. 315--326.

\bibitem{Potdar2014}
A.~Potdar and E.~Shihab, ``An exploratory study on self-admitted technical
  debt,'' in \emph{Proceedings of the 2014 IEEE International Conference on
  Software Maintenance and Evolution}, ser. ICSME '14.\hskip 1em plus 0.5em
  minus 0.4em\relax USA: IEEE Computer Society, 2014, p. 91–100.

\bibitem{Maldonado2015}
E.~da~S.~Maldonado and E.~Shihab, ``Detecting and quantifying different types
  of self-admitted technical debt,'' in \emph{7th {IEEE} International Workshop
  on Managing Technical Debt, {MTD} 2015, Bremen, Germany, October 2, 2015},
  2015, pp. 9--15.

\bibitem{Ren2019TOSEM}
X.~Ren, Z.~Xing, X.~Xia, D.~Lo, X.~Wang, and J.~Grundy, ``Neural network-based
  detection of self-admitted technical debt: From performance to
  explainability,'' \emph{ACM Trans. Softw. Eng. Methodol.}, vol.~28, no.~3,
  jul 2019.

\bibitem{Guo2021}
Z.~Guo, S.~Liu, J.~Liu, Y.~Li, L.~Chen, H.~Lu, and Y.~Zhou, ``{How Far Have We
  Progressed in Identifying Self-admitted Technical Debts? A Comprehensive
  Empirical Study},'' \emph{ACM Transactions on Software Engineering and
  Methodology}, vol.~30, no.~4, pp. 1--56, jul 2021.

\bibitem{Yu2022TSE}
Z.~Yu, F.~M. Fahid, H.~Tu, and T.~Menzies, ``{Identifying Self-Admitted
  Technical Debts With Jitterbug: A Two-Step Approach},'' \emph{IEEE
  Transactions on Software Engineering}, vol.~48, no.~5, pp. 1676--1691, 2022.

\bibitem{ZhouEtAl2019}
Y.~Zhou, S.~Liu, J.~Siow, X.~Du, and Y.~Liu, ``{Devign: Effective vulnerability
  identification by learning comprehensive program semantics via graph neural
  networks},'' in \emph{Advances in Neural Information Processing Systems},
  vol.~32, 2019.

\bibitem{Chakraborty2022}
S.~Chakraborty, R.~Krishna, Y.~Ding, and B.~Ray, ``{Deep Learning Based
  Vulnerability Detection: Are We There Yet?}'' \emph{IEEE Transactions on
  Software Engineering}, vol.~48, no.~9, pp. 3280--3296, 2022.

\bibitem{SteenhoekEtAl2023}
B.~Steenhoek, M.~M. Rahman, R.~Jiles, and W.~Le, ``An empirical study of deep
  learning models for vulnerability detection,'' in \emph{2023 IEEE/ACM 45th
  International Conference on Software Engineering (ICSE)}, 2023, pp.
  2237--2248.

\bibitem{FuTantithamthavorn2022}
M.~Fu and C.~Tantithamthavorn, ``Linevul: A transformer-based line-level
  vulnerability prediction,'' in \emph{2022 IEEE/ACM 19th International
  Conference on Mining Software Repositories (MSR)}, 2022, pp. 608--620.

\bibitem{Feng2020}
Z.~Feng, D.~Guo, D.~Tang, N.~Duan, X.~Feng, M.~Gong, L.~Shou, B.~Qin, T.~Liu,
  D.~Jiang, and M.~Zhou, ``{CodeBERT: A pre-trained model for programming and
  natural languages},'' \emph{Findings of the Association for Computational
  Linguistics Findings of ACL: EMNLP 2020}, pp. 1536--1547, 2020.

\bibitem{Fu2022}
M.~Fu and C.~Tantithamthavorn, ``{LineVul: A Transformer-based Line-Level
  Vulnerability Prediction},'' in \emph{Proceedings of the 19th International
  Conference on Mining Software Repositories}.\hskip 1em plus 0.5em minus
  0.4em\relax New York, NY, USA: ACM, may 2022, pp. 608--620.

\bibitem{Aiken2023}
W.~Aiken, P.~K. Mvula, P.~Branco, G.-V. Jourdan, M.~Sabetzadeh, and H.~Viktor,
  ``{Measuring Improvement of F$_1$-Scores in Detection of Self-Admitted
  Technical Debt},'' in \emph{International Conference on Technical Debt 2023
  (TechDebt)}, 2023.

\bibitem{MockEtAl2024Dataset}
M.~Mock, J.~Melegati, M.~Kretschmann, N.~E.~D. Ferreyra, and B.~Russo,
  ``Made-wic: Multiple annotated datasets for exploring weaknesses in code,''
  in \emph{39th IEEE/ACM International Conference on Automated Software
  Engineering (ASE '24), October 27-November 1, 2024, Sacramento, CA, USA},
  2024.

\bibitem{Fan2020}
J.~Fan, Y.~Li, S.~Wang, and T.~N. Nguyen, ``A c/c++ code vulnerability dataset
  with code changes and cve summaries,'' in \emph{Proceedings - 2020 IEEE/ACM
  17th International Conference on Mining Software Repositories, MSR
  2020}.\hskip 1em plus 0.5em minus 0.4em\relax Association for Computing
  Machinery, Inc, 2020, pp. 508--512.

\bibitem{YangEtAl2023}
X.~Yang, S.~Wang, Y.~Li, and S.~Wang, ``Does data sampling improve deep
  learning-based vulnerability detection? yeas! and nays!'' in \emph{2023
  IEEE/ACM 45th International Conference on Software Engineering (ICSE)}, 2023,
  pp. 2287--2298.

\bibitem{Russell2019}
R.~Russell, L.~Kim, L.~Hamilton, T.~Lazovich, J.~Harer, O.~Ozdemir,
  P.~Ellingwood, and M.~McConley, ``{Automated Vulnerability Detection in
  Source Code Using Deep Representation Learning},'' \emph{Proceedings - 17th
  IEEE International Conference on Machine Learning and Applications, ICMLA
  2018}, pp. 757--762, 2019.

\bibitem{HarerEtAl2018}
J.~A. Harer, L.~Y. Kim, R.~L. Russell, O.~Ozdemir, L.~Kosta, A.~Rangamani,
  L.~H. Hamilton, G.~I. Centeno, J.~R. Key, P.~M. Ellingwood, M.~W. McConley,
  J.~M. Opper, P.~Chin, and T.~Lazovich, ``Automated software vulnerability
  detection with machine learning,'' \emph{ArXiv}, vol. abs/1803.04497, 2018.

\bibitem{LiEtAl2021}
G.~Lin, J.~Zhang, W.~Luo, L.~Pan, O.~D. Vel, P.~Montague, and Y.~Xiang,
  ``Software vulnerability discovery via learning multi-domain knowledge
  bases,'' \emph{IEEE Transactions on Dependable and Secure Computing},
  vol.~18, no.~05, pp. 2469--2485, sep 2021.

\bibitem{Lin2020}
G.~Lin, W.~Xiao, J.~Zhang, and Y.~Xiang, ``Deep learning-based vulnerable
  function detection: A benchmark,'' in \emph{Information and Communications
  Security}, J.~Zhou, X.~Luo, Q.~Shen, and Z.~Xu, Eds.\hskip 1em plus 0.5em
  minus 0.4em\relax Cham: Springer International Publishing, 2020, pp.
  219--232.

\bibitem{Maldonado2017}
E.~d.~S. Maldonado, E.~Shihab, and N.~Tsantalis, ``Using natural language
  processing to automatically detect self-admitted technical debt,'' \emph{IEEE
  Transactions on Software Engineering}, vol.~43, no.~11, pp. 1044--1062, 2017.

\bibitem{BleiholderNaumann2009DataFusion}
\BIBentryALTinterwordspacing
J.~Bleiholder and F.~Naumann, ``Data fusion,'' \emph{ACM Comput. Surv.},
  vol.~41, no.~1, Jan. 2009. [Online]. Available:
  \url{https://doi.org/10.1145/1456650.1456651}
\BIBentrySTDinterwordspacing

\bibitem{SennrichEtAl2016}
\BIBentryALTinterwordspacing
R.~Sennrich, B.~Haddow, and A.~Birch, ``Neural machine translation of rare
  words with subword units,'' in \emph{Proceedings of the 54th Annual Meeting
  of the Association for Computational Linguistics (Volume 1: Long
  Papers)}.\hskip 1em plus 0.5em minus 0.4em\relax Berlin, Germany: Association
  for Computational Linguistics, Aug. 2016, pp. 1715--1725. [Online].
  Available: \url{https://aclanthology.org/P16-1162}
\BIBentrySTDinterwordspacing

\bibitem{Zhou2021}
X.~Zhou, D.~G. Han, and D.~Lo, ``{Assessing Generalizability of CodeBERT},''
  \emph{Proceedings - 2021 IEEE International Conference on Software
  Maintenance and Evolution, ICSME 2021}, pp. 425--436, 2021.

\bibitem{Liu2019}
\BIBentryALTinterwordspacing
Y.~Liu, M.~Ott, N.~Goyal, J.~Du, M.~Joshi, D.~Chen, O.~Levy, M.~Lewis,
  L.~Zettlemoyer, and V.~Stoyanov, ``Roberta: A robustly optimized bert
  pretraining approach,'' 2019. [Online]. Available:
  \url{https://arxiv.org/abs/1907.11692}
\BIBentrySTDinterwordspacing

\bibitem{SunEtAL2019}
M.~Sun, X.~Huang, H.~Ji, Z.~Liu, and Y.~Liu, Eds., \emph{How to Fine-Tune BERT
  for Text Classification?}\hskip 1em plus 0.5em minus 0.4em\relax Cham:
  Springer International Publishing, 2019.

\bibitem{RussoEtAl2024Replication}
\BIBentryALTinterwordspacing
B.~Russo, J.~Melegati, and M.~Mock, ``Replication package: Leveraging
  multi-task machine learning to improve vulnerability detection.'' [Online].
  Available:
  \url{https://github.com/moritzmock/multitask-vulberability-detection}
\BIBentrySTDinterwordspacing

\bibitem{Ghosh2024}
K.~Ghosh, C.~Bellinger, R.~Corizzo, P.~Branco, B.~Krawczyk, and N.~Japkowicz,
  ``{The class imbalance problem in deep learning},'' \emph{Machine Learning},
  vol. 113, no.~7, pp. 4845--4901, 2024.

\bibitem{Krawczyk2016}
B.~Krawczyk, ``{Learning from imbalanced data: open challenges and future
  directions},'' \emph{Progress in Artificial Intelligence}, vol.~5, no.~4, pp.
  221--232, nov 2016.

\bibitem{Alves2014}
N.~S.~R. Alves, L.~F. Ribeiro, V.~Caires, T.~S. Mendes, and R.~O. Sp\'{\i}nola,
  ``Towards an ontology of terms on technical debt,'' in \emph{2014 Sixth
  International Workshop on Managing Technical Debt}, 2014, pp. 1--7.

\bibitem{Huang2018}
Q.~Huang, E.~Shihab, X.~Xia, D.~Lo, and S.~Li, ``Identifying self-admitted
  technical debt in open source projects using text mining,'' \emph{Empirical
  Software Engineering}, vol.~23, no.~1, pp. 418--451, Feb 2018.

\bibitem{Li2021FSE}
Y.~Li, S.~Wang, and T.~N. Nguyen, ``Vulnerability detection with fine-grained
  interpretations,'' in \emph{Proceedings of the 29th ACM Joint Meeting on
  European Software Engineering Conference and Symposium on the Foundations of
  Software Engineering}.\hskip 1em plus 0.5em minus 0.4em\relax New York, NY,
  USA: Association for Computing Machinery, 2021, p. 292–303.

\bibitem{Xavier2020}
L.~Xavier, F.~Ferreira, R.~Brito, and M.~T. Valente, ``{Beyond the Code: Mining
  Self-Admitted Technical Debt in Issue Tracker Systems},'' in
  \emph{Proceedings of the 17th International Conference on Mining Software
  Repositories}.\hskip 1em plus 0.5em minus 0.4em\relax New York, NY, USA: ACM,
  jun 2020, pp. 137--146.

\bibitem{Li2022ese}
Y.~Li, M.~Soliman, and P.~Avgeriou, ``{Identifying self-admitted technical debt
  in issue tracking systems using machine learning},'' \emph{Empirical Software
  Engineering}, vol.~27, no.~6, pp. 1--37, 2022.

\bibitem{Sierra2019}
G.~Sierra, E.~Shihab, and Y.~Kamei, ``A survey of self-admitted technical
  debt,'' \emph{Journal of Systems and Software}, vol. 152, pp. 70--82, 2019.

\bibitem{Vaswani2002}
A.~Vaswani, ``{Attention is all you need},'' \emph{IEEE Industry Applications
  Magazine}, vol.~8, no.~1, pp. 8--15, 2002.

\bibitem{Wang2020}
X.~Wang, J.~Liu, L.~Li, X.~Chen, X.~Liu, and H.~Wu, ``{Detecting and explaining
  self-admitted technical debts with attention-based neural networks},'' in
  \emph{Proceedings of the 35th IEEE/ACM International Conference on Automated
  Software Engineering}.\hskip 1em plus 0.5em minus 0.4em\relax New York, NY,
  USA: ACM, dec 2020, pp. 871--882.

\bibitem{Tu2022}
H.~Tu and T.~Menzies, ``{DebtFree: minimizing labeling cost in self-admitted
  technical debt identification using semi-supervised learning},''
  \emph{Empirical Software Engineering}, vol.~27, no.~4, 2022.

\bibitem{Nam2016}
J.~Nam and S.~Kim, ``{CLAMI: Defect prediction on unlabeled datasets},''
  \emph{Proceedings - 2015 30th IEEE/ACM International Conference on Automated
  Software Engineering, ASE 2015}, pp. 452--463, 2016.

\end{thebibliography}

\end{document}